\documentclass{aa}
\usepackage{txfonts}
\usepackage{graphicx}

\begin{document}

\title{The weak \textit{INTEGRAL} bursts \object{GRB\,040223} and \object{GRB\,040624}: an emerging population  of dark afterglows\thanks{Based on observations made by \textit{XMM-Newton} and with ESO telescopes under programmes Id 72.D-0480 and 073.D-0255 and with the TNG. \textit{INTEGRAL} is an ESA project with instruments
and science data centre funded by ESA member states (especially the PI
countries: Denmark, France, Germany, Italy, Switzerland, Spain), Czech Republic, and Poland, and with the participation of Russia and the USA. }}

 \author{P. Filliatre\inst{1,2} \and S. Covino\inst{3} \and P. D'Avanzo\inst{3,4} \and A. De Luca\inst{5} \and D. G\"{o}tz\inst{5} \and S. McGlynn\inst{6} \and S. McBreen\inst{7} \and  D. Fugazza\inst{3}  \and A. Antonelli\inst{8} \and S. Campana\inst{3}  \and G. Chincarini\inst{3,9} \and A. Cucchiara\inst{3} \and M. Della Valle\inst{10} \and S. Foley\inst{6} \and P. Goldoni\inst{1,2}  \and L. Hanlon\inst{6} \and G. Israel\inst{8} \and B. McBreen\inst{6} \and S. Mereghetti\inst{5} \and L. Stella\inst{8} \and G. Tagliaferri\inst{3} }

\offprints{filliatr@cea.fr}

\institute{ 
Laboratoire Astroparticule et Cosmologie, UMR 7164, 11 place Marcelin Berthelot, F-75231 Paris Cedex 05, France
\and  
Service d'Astrophysique, CEA/DSM/DAPNIA/SAp, CE-Saclay, Orme des Merisiers, B\^{a}t. 709, F-91191 Gif-sur-Yvette Cedex, France
\and        
INAF, Osservatorio Astronomico di Brera, via E. Bianchi 46, I-23807 Merate (LC), Italy
\and 
Universita` Insubria, Dipartimento di Fisica e Matematica, via Valleggio 11, 22100, Como, Italy
\and 
INAF - Istituto di Astrofisica Spaziale e Fisica Cosmica di Milano, via E. Bassini 15, I-20133 Milano, Italy
\and 
School of Physics, University College Dublin, Dublin 4, Ireland
\and 
Astrophysics Missions Division,
Research Scientific Support Department of ESA, ESTEC, Noordwijk,
The Netherlands
\and 
INAF, Osservatorio Astronomico di Roma, via Frascati 33, Monteporzio Catone, 00040 Rome, Italy
\and 
Universit\`a degli Studi di Milano-Bicocca, piazza dell'Ateneo Nuovo 1, I-20126 Milano, Italy.
\and 
INAF, Osservatorio Astrofisico di Arcetri, largo E. Fermi 5, 50125 Firenze, Italy
}

 \date{Received <date> / Accepted <date>}

\titlerunning{The dark \object{GRB\,040223} and \object{GRB\,040624}}
\authorrunning{P. Filliatre et al.}

 \abstract{We report here $\gamma$-ray, X-ray and near-infrared observations of \object{GRB\,040223} along with
 $\gamma$-ray and optical observations of \object{GRB\,040624}.
 \object{GRB\,040223} was detected by \textit{INTEGRAL} close to the Galactic plane and \object{GRB\,040624} at high
 Galactic latitude. Analyses of the prompt emission detected by the IBIS instrument on \textit{INTEGRAL}
 are presented for both bursts. The two GRBs have long durations, slow pulses and
are weak. The $\gamma$-ray spectra of both bursts are
 best fit with steep power-laws, implying they are X-ray rich. \object{GRB\,040223} is among the weakest and longest of
 \textit{INTEGRAL} GRBs.
The X-ray afterglow of this burst was detected 10 hours after
the prompt event by \textit{XMM-Newton}. The measured spectral
properties are consistent with a  column density much higher
than that expected from the Galaxy, indicating strong
intrinsic absorption. We carried out near-infrared observations 17
hours after the burst with the NTT of ESO, which yielded upper
limits. Given the intrinsic absorption, we find that these limits
are compatible with a simple extrapolation of the X-ray afterglow
properties. For \object{GRB\,040624}, we carried out optical
observations 13 hours after the burst with FORS\,1 and 2 at the
VLT, and DOLoRes at the TNG, again obtaining upper limits.
We compare these limits with the magnitudes of a compilation
of promptly observed counterparts of previous GRBs and show that
they lie at the very faint end of the distribution. These
two bursts are good examples of a population of bursts with dark
or faint afterglows that are being unveiled through the
increasing usage of large diameter telescopes engaged in
comprehensive observational programmes.
 }

\maketitle

%

\section{Introduction}
\label{sc:intro}

The emergence of comprehensive observational follow-up
programmes for GRBs means that it is possible to have, for a
particular GRB, a set of multiwavelength observations going from
the $\gamma$-ray to the radio domain, conveying a host of
information about the nature of these events (\cite{zhang}).
Moreover, the GRB afterglows are a promising tool for cosmology,
as their absorption spectra lead to the determination of the
redshift and the study of the chemical environment of a new set of
galaxies (e.g. \cite{fiore}), with the possibility of
exploration up to the reionisation epoch (\cite{lamb,cus,tag}).\\
\indent However, before they are used for cosmological purposes, it is critical to explore the inhomogeneities among the GRBs, in particular to establish if  there are different classes of afterglow
behaviour. One such classification distinguishes between bursts
with  (`bright') and without (`dark') detected optical
afterglows.\\
\indent The estimated fraction of GRBs which did not show any detectable
 afterglow in the optical band depends strongly on the detection satellite: from less than 10\% for
 HETE II bursts (\cite{hete}), to 60\% for events detected by BeppoSAX, XTE or the Interplanetary Network
 (\cite{lazzati}). Popular and non-mutually exclusive explanations are: these bursts have intrinsically faint afterglows
 in the optical band
(e.g. \cite{fynbo,lazzati}); their decay is very fast
(\cite{berger}); the optical afterglow is obscured by dust
in the vicinity of the GRB or in the star-forming region in which
the GRB occurs (e.g. \cite{lamb2000,reichart}); their redshift is
above 5, so that the Lyman-$\alpha$ absorption by neutral hydrogen
in the host galaxy and along the line of
sight suppresses the optical radiation of the afterglow (\cite{lamb2000,tag}).\\
 \indent To these physical explanations, one must add the complication that the search techniques may be neither
 accurate nor quick enough (\cite{hete}). The continuing reduction of observational biases in afterglow searches,
 and the corresponding increased efficiency of observing facilities, often robotic, devoted to this task has therefore
 modified the percentage of GRBs with no identified optical or near-infrared (NIR) counterpart. However, the discrepancy
 between the high-energy (soft X-ray) afterglows and those identified at longer wavelengths (optical and NIR, but also
 radio) still holds true. In Table~\ref{tab:afterglow:search} we give the numbers of detected afterglows, using  the
 publicly available catalogue maintained by J.~Greiner\footnote{\texttt{http://www.mpe.mpg.de/$\sim$jcg/grbgen.html}}
 and \cite{meregotz} for bursts detected by  \textit{INTEGRAL} (\cite{winkler}). In order to determine the real percentage
 of dark afterglows, it is  essential to consider in detail some of the various factors affecting the observers'
 capabilities to single out the counterparts, i.e. the size of the error boxes and the delay in making the position
 available after the high-energy event. Moreover, for \textit{INTEGRAL}, one has to add the fact that this satellite
 preferentially points to the Galactic plane, and therefore tends to select bursts whose afterglow emission is
 heavily obscured at optical wavelengths.\\
\begin{table}
\centering \caption{Numbers of detected afterglows in the X-ray,
optical and radio domains, for a total of 276 GRBs localised
before July 2005 (X-ray flashes are excluded).}
\label{tab:afterglow:search}
\begin{tabular}{lrrrrr}
\hline
                            &  BeppoSAX  & HETE\,II &  \textit{INTEGRAL}  &  Swift & Other\\
\hline
GRBs    &          53 &           55 &           28 &           41 &           99\\
X-ray     &          30 &           10 &            11 &           28 &           10\\
Optical   &          16 &           20 &            9 &           15 &           14\\
Radio     &          10 &            6 &            2 &            5 &           11\\
\hline
\end{tabular}
\end{table}
\indent In this paper, we present and discuss our observations of
two GRBs discovered by \textit{INTEGRAL}, \object{GRB\,040223} and
\object{GRB\,040624}. \object{GRB\,040223} is located in the
direction of the Galactic plane. The X-ray afterglow was
detected with \textit{XMM-Newton}, allowing the error box to
be reduced significantly. The absorption measured in the X-ray
afterglow spectrum is consistent with being about 2.4 times that
due to the Galactic plane. However no afterglow was found in our
deep observations in the $J$, $H$ and $K_{\rm s}$ bands carried out less than 17 hours after the burst.
\object{GRB\,040624} was very well situated at high Galactic
latitude. However, our rather deep observations carried out within 13 hours of the burst provided only an upper limit on
the magnitude of the afterglow in the optical. There are no X-ray
observations of the afterglow of \object{GRB\,040624}. One day
after the GRB (Feb 24.58), radio observations were performed
 with the Very Large Array (VLA) providing
2-$\sigma$ upper limits of 200 and 174 $\mu$Jy at 4.9 and 8.5 GHz,
respectively (\cite{Soder04}). We present the
$\gamma$-ray characteristics of these two dark bursts in this
paper. In addition, we present
their afterglow properties and compare them to the afterglow population as a whole.\\
\indent The errors quoted are at the 1-$\sigma$ level, unless explicitly stated otherwise.

\section{Observations and data analysis}\label{sc:obs}

\subsection{\object{GRB\,040223}}

\subsubsection{$\gamma$-ray data}
\label{040223gamma}

\begin{figure}
\centering
\resizebox{\hsize}{!}{\includegraphics[angle=270,width=\columnwidth]{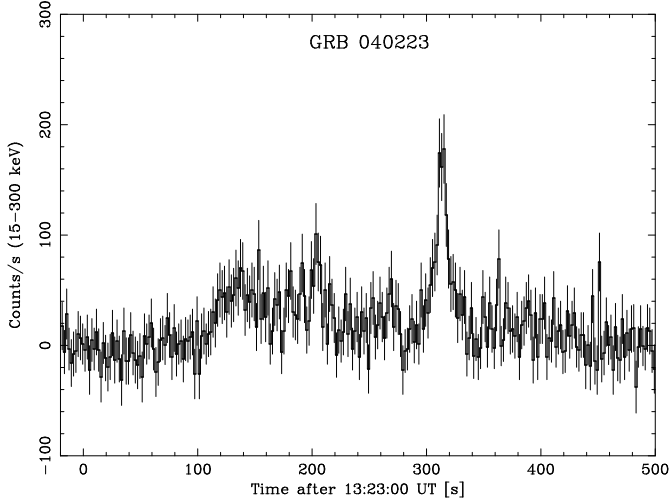}}
\caption{IBIS/ISGRI light curve of \object{GRB\,040223} in the
15-300 keV band in vignetting corrected counts.} \label{040223.lc}
\end{figure}

\begin{figure}
\centering
\resizebox{\hsize}{!}{\includegraphics[angle=270,width=\columnwidth]{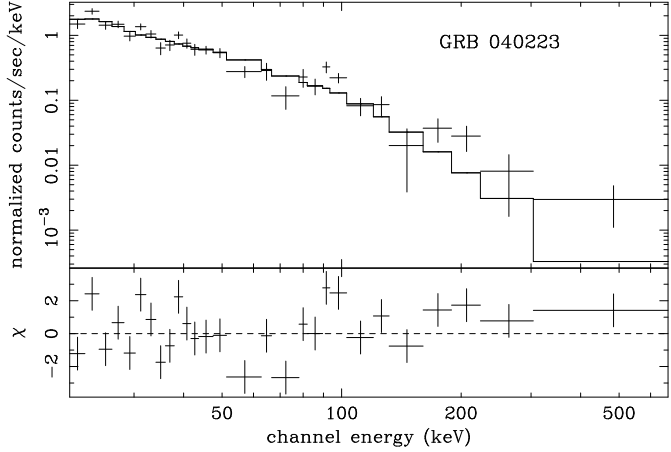}}
\caption{IBIS/ISGRI average spectrum of \object{GRB\,040223} fit with a power-law model.} \label{040223sp}
\end{figure}

\object{GRB\,040223} was detected with IBIS/ISGRI (Imager on Board
the \textit{INTEGRAL} Satellite, \cite{uber2003}) at 13:28:10 UTC
(\cite{Gotz04a}), and localised thanks to the \textit{INTEGRAL} Burst
  Alert System, IBAS (\cite{Mere03,mere:2004}). The IBAS alert was issued at 13:28:30, 30 seconds
after the beginning of the main peak and 3 minutes and 50 s
after the onset of the weak precursor emission. The
coordinates are $\alpha=16^{\rm h}39^{\rm m}31^{\rm s}$, $\delta =
-41\degr55\arcmin47\arcsec$ (J2000), with an uncertainty of
$2\farcm2$ (90\,\% confidence limit, hereafter CL).
The Galactic coordinates are  $l=341\fdg61$, $b=3\fdg19$. \\
\indent The light curve (15-300 keV) is given in
Fig.~\ref{040223.lc}. The GRB starts at 13:24:40 UT and has a
 $T_{90}$ duration of $258\,\rm s$, making it one of the
longest GRBs seen by \textit{INTEGRAL}. It exhibits a
multi-peaked structure, with a main peak preceded by two
faint precursors.\\
\indent The peak spectrum (integrated over 1 s) and the
average spectrum have been extracted for the burst and fit
using the latest available response matrices. In the 20-200 keV
energy range, the peak flux and fluence are $4\times10^{-8}\,\rm
erg\,cm^{-2}\,s^{-1}$ and $2\times10^{-6}\,\rm erg\,cm^{-2}$
respectively. \\
The peak spectrum has a photon index of $1.4\pm0.6$, while the
photon index for the entire last peak is $1.82\pm0.17$. The
average spectrum, shown in Fig.~\ref{040223sp}, can be modelled as
a power-law, yielding a photon index of $\Gamma = 2.0\pm0.2$
(90\,\% CL). The results obtained with the germanium spectrometer
SPI on INTEGRAL (\cite{ved2003}) are consistent with IBIS 
(\cite{mcglynn}).


\subsubsection{The X-ray afterglow}

\begin{figure}
\centering
\resizebox{\hsize}{!}{\includegraphics[angle=270,width=\columnwidth]{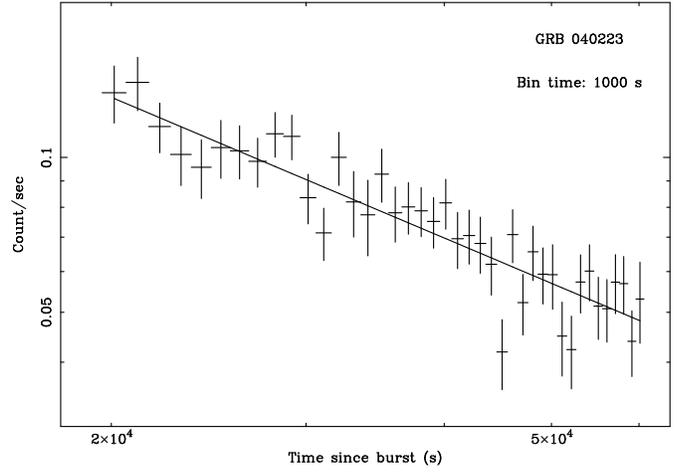}}
\caption{Background-subtracted light curve of the afterglow 
of \object{GRB\,040223} observed by the EPIC instrument (1-7
keV). The decay is well fit by a power-law $F\propto
t^{-\alpha_{\rm X}}$, with best fit value of $\alpha_{\rm
X}=0.9\pm0.1$ (90\% CL, $\chi^2_{\nu}=1.05$, 39 d.o.f.).}
\label{grb040223:lc:x}
\end{figure}

The \textit{INTEGRAL} error box was observed by
\textit{XMM-Newton} on February 23, 2004 at 18:21 UTC, slightly
less than 5 hours after the burst. This is  the fastest response
by XMM-Newton to a GRB trigger to date. The
observation lasted for $\sim42$ ks. We report here on the data
collected by the EPIC instrument, consisting of the PN and
two MOS cameras
(\cite{strueder01,turner01}). \\
\indent The afterglow of \object{GRB\,040223} (source XMMU
J163929.9-415601) is clearly detected in all cameras
(\cite{Breit04,DeL04}). We improved the EPIC astrometry by
cross-correlating the serendipitous X-ray sources in the field
with stars in the USNO-B1 catalog. The refined coordinates for the
X-ray afterglow are $\alpha=16^{\rm h}39^{\rm m}30\fs17$,
$\delta=-41\degr55\arcmin59\farcs7$ with an error of 
$1\farcs5$, accounting for the accuracy of the X-ray to optical
source superposition, as well as for the positional accuracy of
the USNO-B1 catalog.
The afterglow faded during the observation, as can be seen
 in Fig.~\ref{grb040223:lc:x}. After checking the consistency
of the results from the three detectors, we extracted a combined
background-subtracted light curve in the range 1-7 keV. The count
rate decay is well described ($\chi^2_{\nu}=1.05$, 39 d.o.f.) by a
power-law ($F(t)\propto t^{-\alpha_{\rm X}}$) with the
temporal index $\alpha_{\rm X}=0.9\pm0.1$ (90\% confidence level
for a single
parameter of interest). \\
\begin{figure}
\centering
\resizebox{\hsize}{!}{\includegraphics[angle=270,width=\columnwidth]{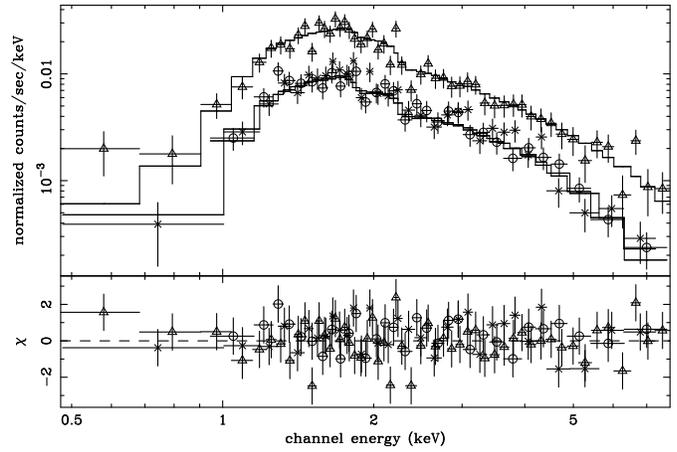}}
\caption{EPIC spectra of the afterglow of
\object{GRB\,040223}, overlaid with the best fit model
convolved through the instrument response matrices and effective
area files. Triangles, circles and crosses represent PN, MOS1 and
MOS2 data, respectively.} \label{grb040223:sp:x}
\end{figure}
\indent Source and background spectra were extracted. The spectral
analysis was performed using XSPEC v11.3. The spectra from the
three detectors, shown in Fig.~\ref{grb040223:sp:x}, were 
fit simultaneously. A simple absorbed power-law model
yields a good description of the data ($\chi^2_{\nu}=0.93$, 114
d.o.f.). The best fit values are: $2.60\pm0.15$ (90\%  CL)
for the photon index, $N_{\rm H}=(1.60\pm0.15)\times10^{22}\,\rm
cm^{-2}$ for the hydrogen column density. This is 
significantly ($3.7\sigma$) higher than the expected contribution
of the Galaxy, which is estimated to be $N_{\rm
H}=(6.6\pm2)\times10^{21}\,\rm cm^{-2}$ (\cite{dickey}), where the
face value and the error are the average and the standard
deviation respectively of the 7 nearest available
measurements within $1\degr$ of the GRB position. This
implies an unabsorbed flux of $\sim1.0\times10^{-12}\,\rm
erg\,cm^{-2}\,s^{-1}$in the band 0.5-10 keV. These results, fully
consistent with those found by \cite{mcglynn}, still hold when two
subsets of the data with the same number of counts (having
exposure times of $\sim15$ ks and $\sim27$ ks, respectively) 
are considered separately, showing that there is no significant
variation in the spectral parameters with time.\\
\indent We investigated the presence of absorption or emission
lines superimposed on the continuum. We added Gaussian lines of
fixed width (smaller than the instrumental energy resolution),
allowing the central energy in the range 1-5 keV to vary, as
well as the line normalisation (in both the emission and
absorption cases). We found no significant features. The upper
limits (3-$\sigma$) on the equivalent width of any emission or
absorption lines are $\sim60\,\rm eV$ and $\sim200\,\rm eV$ in the
1-2.5 keV and 2.5-5 keV ranges, respectively.

\subsubsection{Optical and near-infrared observations}\label{subopt:040223}

\begin{table*}
\centering \caption{Observation log for \object{GRB\,040223}.
Observing times refer to the middle of the exposures 
and $\Delta T$ is the time since the GRB. Magnitudes are
calibrated following the 2\,MASS catalogue. Magnitude limits are
at the 3-$\sigma$ level.} \label{tab:obslog040223}
\begin{tabular}{llcrcll}
\hline
\bf Date         & $\Delta$ T         &\bf Filter &\bf Exp   &\bf Seeing  &\bf Instrument & \bf Mag\\
(UT)             &  (days)            &                    &\multicolumn{1}{c}{(s)}  &(\arcsec)   &  &\\
\hline
2004 Feb 24.279 &  0.720   & $J$               & 20 sec     & 1.2        & NTT+SofI  & $>20.9$ \\
2004 Feb 24.258  &  0.700   & $H$               & 20 sec     & 1.1        & NTT+SofI   & $>19.8$ \\
2004 Feb 24.259 &  0.700   & $K_{\rm s}$              & 10 sec     & 1.1        & NTT+SofI   & $>19.9$ \\
2004 Feb 24.384 &   0.824   & $J$               & 20 sec     & 0.687      & NTT+SofI   & $>21.1$ \\
2004 Feb 24.394 &   0.836   & $H$               & 20 sec     & 0.67       & NTT+SofI  & $>20.4$ \\
2004 Feb 24.403 &   0.844   &   $K_{\rm s}$             & 10 sec     & 0.638      & NTT+SofI   & $>20.2$ \\
2004 Feb 24.538 &   1.821    & $K_{\rm s}$               & 10 sec     & 0.75       & NTT+SofI    & $>20.1$ \\
\hline
\end{tabular}
\end{table*}

\begin{figure}
\resizebox{\hsize}{!}{\includegraphics[width=\columnwidth]{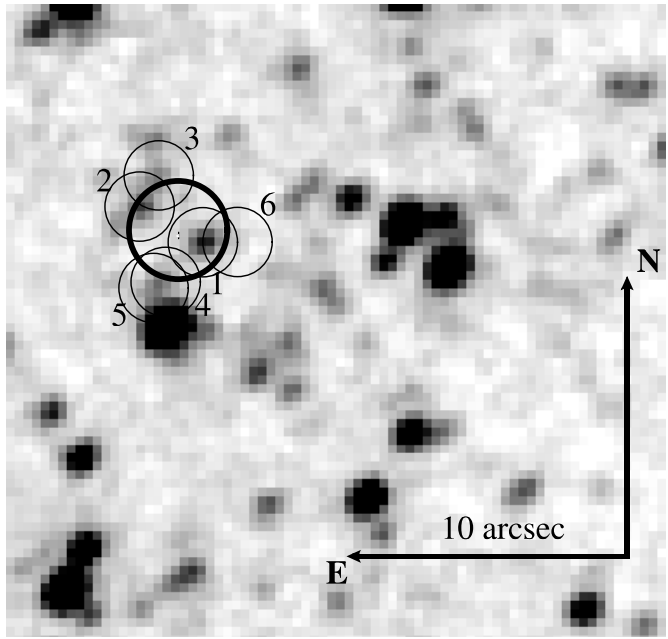}}
\caption{Finding chart for \object{GRB\,040223}, with 
the \textit{XMM-Newton} error circle shown. Six
objects are singled out within the 1-$\sigma$ error circle, but
none of these show the typical afterglow variability.}
\label{fig:XMM_040223}
\end{figure}

The first follow-up observations were carried out by the
Faulkes Telescope North  1.57 hours after the GRB in the optical,
obtaining a 2-$\sigma$ upper limit of $r'\sim18$ 
(\cite{Gomb04}). The GRB was within the Galactic
plane, not very far from the Galactic centre, so NIR
observations are better suited to overcome the high
absorption. The REM robotic telescope (\cite{Zerbi01,chinc})
provided NIR  upper limits (2-$\sigma$) of $H=15.5$ and
$K=15.0$
16 hours after the GRB trigger (\cite{Isra04}). \\
\indent A much deeper search for the afterglow in the NIR
 was carried out with a similar delay using the SofI instrument at the NTT (\cite{Tagl04,Simon04}) . We
made three sets of observations, from 0.7 to 1.8 days after
the burst (Table~\ref{tab:obslog040223}). Image reduction
was performed following the standard procedures of the ESO -
Eclipse package (\cite{devillard}). Astrometry and
photometric calibration were performed using the 2\,MASS catalogue\footnote{\textit{http://irsa.ipac.caltech.edu/}}.
PSF-photometry of all the objects in the field was
carried out using the \textit{IRAF}\footnote{\textit{IRAF} is
distributed by the National Optical Astronomy Observatories,
    which are operated by the Association of Universities for Research
    in Astronomy, Inc. under cooperative agreement with the National
    Science Foundation.} daophot task.
A finding chart of the field, with
the reported \textit{XMM-Newton} error box, is shown in Fig.~\ref{fig:XMM_040223}.\\
\indent Within the \textit{XMM-Newton} error circle of
$1\farcs5$ radius we found 6 objects, for which magnitude
measurements were carried out in the $K_{\rm s}$ band at
$\Delta T=0.70$, 0.83 and 1.8 days. For each of them, we fit a
power-law decay over the three epochs of observation in the
$K_{\rm s}$ band and found that a temporal index consistent
with zero is always preferred, indicating no source
variability. Fixing $\alpha=\alpha_{\rm X}=0.9\pm0.1$ leads to
unacceptable $\chi^2$ values. Therefore, the limiting magnitudes reported in
Table~\ref{tab:obslog040223} are our 3-$\sigma$ upper limits on the magnitude of the afterglow.\\
\indent These magnitude limits are not corrected for
Galactic absorption. The afterglow lies towards the Galactic
plane, rather close to the direction of the Galactic centre. The
extinction is high, rather patchy, and uncertain close to the
Galactic centre. Using the previously quoted value of $N_{\rm
H}=(6.6\pm2)\times10^{21}\,\rm cm^{-2}$, and the fit of
\cite{predehl}, we derive $A_{V}=3.9$; and hence, using the
extinction law of \cite{cardelli} with $R_{V}=3.1$, we get
$E(B-V)_{N_{\rm H}}=1.2$.  The use of the FIRAS dust emission maps
at $100\,\mu\rm m$ obtained by \cite{schlegel} leads to a
substantially higher value of $E(B-V)_{\rm FIR}=1.8$. For this
overestimate of the reddening, typical for low Galactic latitudes,
\cite{dutra} proposed a linear correction. In our case, we then
get $E(B-V)=1.2$, in full agreement with the value derived from
the hydrogen column density. We then derived  $A_R=3.2$,
$A_J=1.1$, $A_H=0.7$ and $A_K=0.4$ for the Galactic
absorption, and we will adopt these values. It should be
noted that the value of $N_{\rm H}$ derived from the spectrum of
the X-ray afterglow is much higher than the one derived for the
Galaxy. This is discussed in Sec.~\ref{sec:discuss}.

\subsection{\object{GRB\,040624}}

\subsubsection{$\gamma$-ray data}

\begin{figure}
\centering
\resizebox{\hsize}{!}{\includegraphics[width=\columnwidth,angle=270]{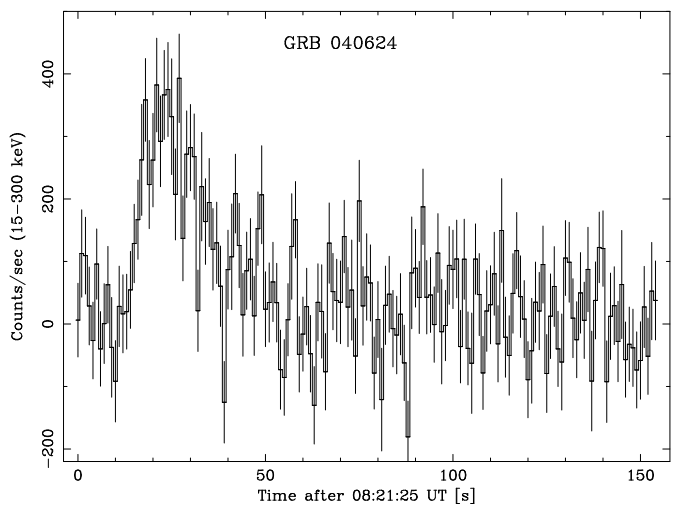}}
\caption{Light curve of \object{GRB\,040624} in the 15-300 keV
band.} \label{040624_lc}
\end{figure}

\begin{figure}
\centering
\resizebox{\hsize}{!}{\includegraphics[width=\columnwidth]{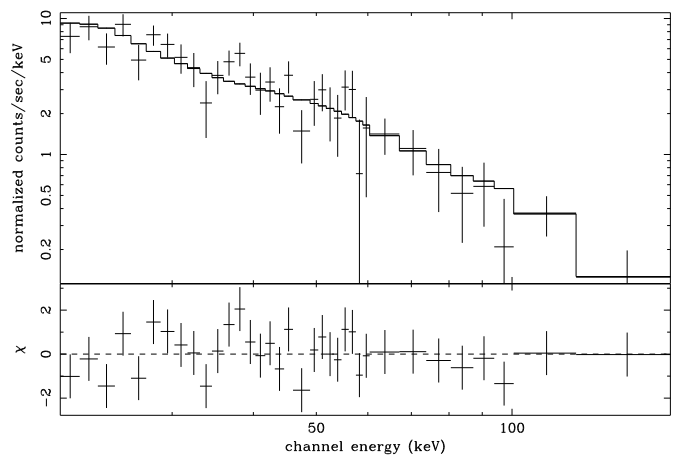}}
\caption{IBIS/ISGRI average spectrum of \object{GRB\,040624} fit by a
power-law model.} \label{040624_spec}
\end{figure}

GRB\,040624 was detected on June 24 at 08:21:35 UT (June 24.348)
with IBIS/ISGRI in the 15-200\,keV band (\cite{Mere04b}). It
was localised with IBAS at $\alpha=13^{\rm h}00^{\rm
m}09\fs9$, $\delta=-03\degr35\arcmin14\arcsec$ with an uncertainty
of 3\arcmin (90\% CL), at $59.2\degr$ from the
Galactic plane. Data from SPI are unavailable for this burst
because the instrument was being annealed. \\
\indent The IBIS data were analysed using standard $\textit
INTEGRAL$ software in a manner similar to
\object{GRB\,040223} (Sec. \ref{040223gamma}).
 The light curve in the energy range 15-300~keV is given in Fig.~\ref{040624_lc}
 and shows the burst consists of one long, slow pulse. The duration is $T_{90}=46\,\rm s$.\\
 \indent  The spectrum obtained for \object{GRB\,040624} is shown in Fig.~\ref{040624_spec}. Each bin contains
 a minimum of  20 counts per bin. The peak flux (integrated over one second) in the energy range 20-200~keV is
  $4\times10^{-8}\,\rm erg\,cm^{-2}\,s^{-1}$. The fluence in the same energy range is
  $1.0\times10^{-6}\,\rm erg\,cm^{-2}$. The spectrum is best fit by a power-law with  photon index
  $\Gamma= 2.1\pm0.2$.


\subsubsection{Optical observations}

\begin{table*}
\centering\caption{Observation log for GRB\,040624. Observing
dates are referred to the middle of the exposures. Magnitude
limits are at the 3-$\sigma$ level.}
\label{tab:obslog040624}
\begin{tabular}{llcrcll}
\hline
\bf Date         & $\Delta$ T  &\bf Filter  &\bf Exp                 &\bf Seeing  &\bf Instrument & \bf Mag\\
(UT)             & (days)      &            &\multicolumn{1}{c}{(s)} &(\arcsec)   &               &         \\
\hline
2004 Jun 24.906  & 0.558       & $R$        & 2$\times$120           & 1.2        & TNG+DOLoRes   & $>23.8$    \\
2004 Jun 24.989  & 0.641       & $R$        & 5$\times$30            & 0.25       & VLT+FORS\,2   & $>22.5$    \\
2004 Jun 25.125  & 0.776       & $R$        & 5$\times$30            & 0.33       & VLT+FORS\,2   & $>24.5$    \\
2004 Jun 27.987  & 3.638       & $R$        & 5$\times$30            & 1.0        & VLT+FORS\,1   & $>24.3$    \\
2004 Jun 28.889  & 4.541       & $R$        & 4$\times$120           & 2.0        & TNG+DOLoRes   & $>22.9$    \\
2004 Jul 07.422  &13.074       & $R$        & 3$\times$180           & 1.4        & TNG+DOLoRes   & $>24.0$    \\
2004 Jul 14.919  &20.571       & $R$        & 2$\times$180           & 1.3        & TNG+DOLoRes   & $>23.4$    \\
\hline
\end{tabular}
\end{table*}

The first follow-up observations were carried out with the the
152 cm telescope of Bologna University, yielding a magnitude limit
of $R_{\rm c}=20.5$ at  11.8 hours (\cite{Picc04}), and by
the 1.5 m Observatorio de Sierra Nevada (OSN) telescope, yielding
a magnitude limit of $R\sim21$ at 12.5 hours
(\cite{Goros04}). { The GRB was located far from the Galactic
plane at $l=307\fdg19$, $b=59\fdg21$, where the optical extinction
is negligible. We estimate $A_V=0.8$ and $A_R=0.06$ (see
subsection~\ref{subopt:040223})}. Our group followed the field in
the optical domain with DOLoRes at TNG, and FORS\,1/2 at the
 VLT as reported in Table~\ref{tab:obslog040624}, with time delays ranging from 0.553 to 20.571 days
 (\cite{Fugaz04,Davan04}).
 The late time observations had the possibility of detecting an underlying supernova.

\section{Discussion}\label{sec:discuss}

\subsection{The prompt event}

\begin{figure}
\centering
\resizebox{\hsize}{!}{\includegraphics[width=\columnwidth]{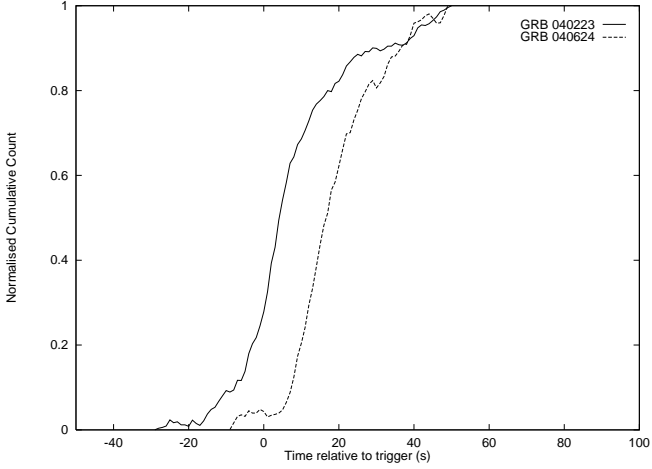}}
    \caption{Normalised cumulative light curves of \object{GRB\,040223} (solid line) and GRB~040624 (dashed line).}
    \label{040624_cumul}
\end{figure}

\begin{figure}
\centering
\resizebox{\hsize}{!}{\includegraphics[angle=270,width=\columnwidth]{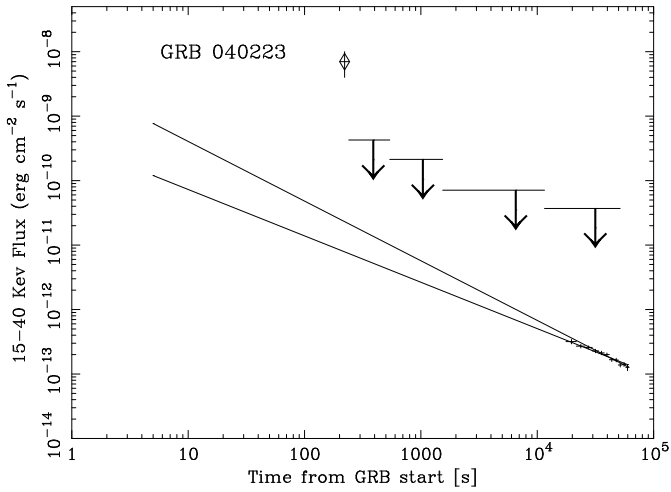}}
\caption{Prompt emission of \object{GRB\,040223} measured by
\textit{INTEGRAL} (diamond) compared with the extrapolation to
earlier times of the X-ray afterglow emission detected by
XMM-Newton within the limits of the measurements (two solid
lines). The arrows indicate IBIS/ISGRI 5-$\sigma$ upper limits for
the emission in the direction of the GRB.} \label{040223ul}
\end{figure}

The two bursts, \object{GRB\,040223} and
\object{GRB\,060624}, are rather faint and about 2/3 of
\textit{INTEGRAL} GRBs have larger peak fluxes
(\cite{meregotz}). The steep power-law spectra (of
$\Gamma=2.0\pm0.2$ for \object{GRB\,040223} and $\Gamma=2.1\pm0.2$
for \object{GRB\,040624}) indicate that they are most likely
X-ray rich GRBs (\cite{barraud}). Indeed, extrapolating this steep
power-law spectrum, we get the flux ratio $S_{7-30\,\rm
keV}/S_{30-400\,\rm keV} \sim 0.5$ for \object{GRB\,040223},
while for classical GRBs this value is smaller than 0.3. \\
\indent Using a large sample of bursts, \cite{mmhq02} {showed}
that the cumulative light curves of GRBs are approximately
linear over the main emission period of the burst, indicating an
almost constant energy output by the central engine. The
normalised cumulative profiles of \object{GRB\,040223} and
\object{GRB\,040624} for the main emission periods, presented in Fig.~\ref{040624_cumul},
exhibit similar profiles showing that most of the
emission evolves linearly with time. \\
\indent The temporal properties of the pulses in GRBs have been
determined and a relationship found between the pulse
properties, the number of pulses and $T_{90}$ 
(\cite{quilligan:2002,mmqh02}. The main pulse of both
\object{GRB\,040223} and \object{GRB\,040624} have slow rise and
fall times, and their pulse properties, along with their
$T_{90}$ durations are quite consistent with the pulse
timing diagrams obtained by \cite{mmqh02}. \object{GRB\,040106} is
another example of an \textit{INTEGRAL} burst that fits
on the timing diagram ({\cite{moran2004}).\\
\indent In many cases the location of a GRB discovered with
IBAS can be monitored for a few hours to days after and
before the burst, thanks to the large FOV of \textit{INTEGRAL} and
its typically long observation time. For
\object{GRB\,040223}, we have performed a search for precursors
and early $\gamma$-ray afterglow in IBIS/ISGRI data using the
Off-line Scientific Analysis (OSA). We have not been able to
detect any precursors or early high-energy afterglow emission.
This burst is, however, much weaker than those for which
high energy afterglows have been reported
(\object{GRB\,920723}: \cite{burenin}, \object{GRB\,940217}:
\cite{hurley}, \object{GRB\,980923}:  \cite{giblin}).\\
\indent Following the work of \cite{frontera00} on BeppoSAX data,
we have compared the X-ray afterglow light curve with the prompt
$\gamma$-ray emission of \object{GRB\,040223} in order to evaluate
the relationship between them. \cite{frontera00} report that the
X-ray afterglow starts at about 50\% of the GRB duration and its
fluence, as computed from the WFC light curve, is consistent with
the decay law found from the afterglow NFI observations. In our
case the extrapolation of the X-ray afterglow is well below the
$\gamma$-ray flux and extrapolates to a time earlier than the
GRB. We should point out that  \cite{tagnat} (see also
\cite{chinc2,nousek}) have shed new light on the
relationship between the GRB and the early X-ray afterglow using
Swift-XRT data. In particular the X--ray afterglows of some
GRBs in
  their sample are characterised by a rapid fall-off ($\alpha_{\rm X}\gse2.5$)
in the first few hundred seconds, followed by a
  less rapid decline ($\alpha_{\rm X}\sim1$}) lasting several hours. The behaviour of \object{GRB\,040223} is
  compatible with this picture and not with that derived from the BeppoSAX GRBs.

\subsection{The optical/NIR afterglow}

\begin{table*}
\centering\caption{A compilation of afterglow data for 39
GRBs. References are given in superscripts.  Delay times
close to our epochs of observation for \object{GRB\,040223} and
\object{GRB\,040624} are preferred.} \label{tab:afterglow}
\begin{tabular}{lllllll}
\hline
\bf Id&\bf Temporal index&\bf Spectral index&$R$&$\Delta T_{R}$ (hours)&$K$&$\Delta T_{K}$ (hours)\\
\hline
\object{GRB\,970228}&$ 1.1\pm 0.1^{\rm (1)}$&$ 0.61\pm 0.32$& 20.83{\rm (2)}& 16.88&&\\
\object{GRB\,971214}&$ 1.20\pm 0.02^{\rm (3)}$&$ 0.93\pm 0.060^{\rm (4)}$& 22.06$^{\rm (3)}$& 12.89& 18.03$^{\rm (5)}$& 3.5\\
\object{GRB\,980326}&$ 2.0\pm 0.1^{\rm (5)}$&$ 0.8\pm 0.4^{\rm (5)}$& 21.25$^{\rm (5)}$& 11.1&&\\
\object{GRB\,980329}&$ 1.21\pm 0.13^{\rm (6)}$&$$& 23.6$^{\rm (6)}$& 19.92&&\\
\object{GRB\,980519}&$ 2.3\pm 0.12^{\rm (7)}$&$ 1.4\pm 0.3^{\rm (7)}$& 20.28$^{\rm (7)}$& 15.59&&\\
\object{GRB\,990123}&$ 1.1\pm 0.03^{\rm (8)}$&$ 0.8\pm 0.1^{\rm (8)}$& 20.0$^{\rm (8)}$& 13.2& 18.29$^{\rm (8)}$& 29.47\\
\object{GRB\,990308}&$ 1.2\pm 0.1^{\rm (9)}$&$$& 18.14$^{\rm (9)}$& 3.34&&\\
\object{GRB\,990510}&$ 0.76\pm 0.01^{\rm (10)}$&$ 0.61\pm 0.12^{\rm (10)}$& 18.9$^{\rm (11)}$& 14.99&&\\
\object{GRB\,990712}&$ 0.97\pm 0.05^{\rm (12)}$&$ 0.7\pm 0.1{\rm (12)}$& 20.175${\rm (12)}$& 10.38&&\\
\object{GRB\,991216}&$ 1.36\pm 0.04^{\rm (13)}$&$ 0.58\pm 0.08^{\rm (13)}$& 19.56$^{\rm (13)}$& 22.5& 15.01$^{\rm (13)}$& 13.16\\
\object{GRB\,000926}&$ 1.69\pm 0.2^{\rm (14)}$&$ 1.42\pm 0.06^{\rm (14)}$& 19.326$^{\rm (14)}$& 20.7&&\\
\object{GRB\,010222}&$ 0.8\pm 0.05^{\rm (15)}$&$ 1.3\pm 0.05^{\rm (15)}$& 18.67$^{\rm (15)}$& 5.85&&\\
\object{GRB\,010921}&$ 1.59\pm 0.18^{\rm (16)}$&$ 2.22\pm 0.23^{\rm (16)}$& 19.4$^{\rm (17)}$& 21.8&&\\
\object{GRB\,011121}&$ 1.72\pm 0.05^{\rm (18)}$&$ 0.66\pm 0.13^{\rm (18)}$& 19.06$^{\rm (18)}$& 10.37&&\\
\object{GRB\,011211}&$ 0.83\pm 0.04^{\rm (19)}$&$ 0.61\pm 0.15^{\rm (19)}$& 20.51$^{\rm (19)}$& 13.01&&\\
\object{GRB\,020124}&$ 1.6\pm 0.04^{\rm (20)}$&$ 1.43\pm 0.14^{\rm (20)}$& 18.376$^{\rm (20)}$& 1.88&&\\
\object{GRB\,020405}&$ 1.54\pm 0.06^{\rm (21)}$&$ 1.3\pm 0.2^{\rm (21)}$& 20.17$^{\rm (21)}$& 23.62&&\\
\object{GRB\,020813}&$ 0.76\pm 0.05^{\rm (22)}$&$ 1.04\pm 0.03^{\rm (22)}$& 18.49$^{\rm (22)}$& 3.97&&\\
\object{GRB\,021211}&$ 1.11\pm 0.01^{\rm (23)}$&$ 0.6\pm 0.2^{\rm (23)}$& 22.2$^{\rm (23)}$& 11.05&&\\
\object{GRB\,030226}&$ 0.50\pm 0.35^{\rm (24)}$&$ 0.7\pm 0.03^{\rm (24)}$& 19.9$^{\rm (24)}$& 15.46& 16.69$^{\rm (24)}$& 8.51\\
\object{GRB\,030328}&$ 1.2\pm 0.1^{\rm (25)}$&$$& 20.91${\rm (25)}$& 15.63&&\\
\object{GRB\,030329}&$ 0.89\pm 0.01^{\rm (26)}$&$ 0.71^{\rm (27)}$& 15.02$^{\rm (27)}$& 9.08&&\\
\object{GRB\,030429}&$ 0.95\pm 0.03^{\rm (28)}$&$ 0.36\pm 0.12^{\rm (28)}$& 20.86$^{\rm (28)}$& 13.15& 17.7$^{\rm (28)}$& 13.63\\
\object{GRB\,030528}&$ 1.2^{\rm (29)}$&$$& 18.7$^{\rm (30)}$& 2.38& 18.6$^{\rm (29)}$& 16.6\\
\object{GRB\,031203}&$ 2.0^{\rm (31)}$&$ 2.36\pm 0.02^{\rm (31)}$&&& 17.56$^{\rm (31)}$& 9.0\\
\object{GRB\,040924}&$ 0.7^{\rm (32)}$&$$& 18.67$^{\rm (33)}$& 6.7& 17.5$^{\rm (34)}$& 2.4\\
\object{GRB\,041006}&$ 1.32\pm 0.02^{\rm (35)}$&$ 0.45^{\rm (36)}$& 20.931$^{\rm (37)}$& 14.94&&\\
\object{GRB\,041219A}&$ 0.8^{\rm (38)}$&$$&&& 16.5$^{\rm (39)}$& 24.24\\
\object{GRB\,041223}&$ 1.1\pm 0.1^{\rm (40)}$&$ 0.6\pm 0.1^{\rm (40)}$& 20.81$^{\rm (41)}$& 14.4&&\\
\object{GRB\,050124}&$ 1.45\pm 0.25^{\rm (40)}$&$ 0.4\pm 0.2^{\rm (40)}$&&& 19.66$^{\rm (41)}$& 24.5\\
\object{GRB\,050315}&$ 0.57^{\rm (42)}$&$$& 20.9$^{\rm (41)}$& 11.6&&\\
\object{GRB\,050319}&$ 0.6^{\rm (43)}$&$$& 20.14$^{\rm (41)}$& 8.7&&\\
\object{GRB\,050401}&$ 1^{\rm (44)}$&$$& 23.18$^{\rm (44)}$& 11.39&&\\
\object{GRB\,050408}&$ 0.68\pm 0.06^{\rm (45)}$&$$& 21.6$^{\rm (45)}$& 8.75&&\\
\object{GRB\,050502A}&$ 1.44\pm 0.2^{\rm (46)}$&$$& 19.1$^{\rm (47)}$& 5.83&&\\
\object{GRB\,050505}&$ 0.7^{\rm (48)}$&$$& 21.5$^{\rm (49)}$& 7.83& 18.1$^{\rm (50)}$& 6.82\\
\object{GRB\,050525A}&$ 1.38^{\rm (51)}$&$$& 18.1$^{\rm (52)}$& 5.2& 16.15$^{\rm (52)}$& 5.2\\
\object{GRB\,050603}&$ 1.97^{\rm (53)}$&$$& 16.5$^{\rm (54)}$& 3.4&&\\
\object{GRB\,050607}&$ 0.5^{\rm (55)}$&$ 1.5^{\rm (55)}$& 22.7$^{\rm (56)}$& 1.27&&\\
\hline
\end{tabular}
\begin{flushleft}
$^{\rm (1)}$\cite{fru},
$^{\rm (2)}$\cite{guarn},
$^{\rm (3)}$\cite{die},
$^{\rm (4)}$\cite{rei},
$^{\rm (5)}$\cite{goro},
$^{\rm (5)}$\cite{blo2},
$^{\rm (6)}$\cite{rei2},
$^{\rm (7)}$\cite{vrba},
$^{\rm (8)}$\cite{kulkarni},
$^{\rm (9)}$\cite{sch},
$^{\rm (10)}$\cite{stanek},
$^{\rm (11)}$\cite{harrison},
$^{\rm (12)}$\cite{sahu},
$^{\rm (13)}$\cite{garna},
$^{\rm (14)}$\cite{fyn},
$^{\rm (15)}$\cite{sta},
$^{\rm (16)}$\cite{pri},
$^{\rm (17)}$\cite{par},
$^{\rm (18)}$\cite{gar},
$^{\rm (19)}$\cite{hol},
$^{\rm (20)}$\cite{berger},
$^{\rm (21)}$\cite{mas2},
$^{\rm (22)}$\cite{cov},
$^{\rm (23)}$\cite{pandey},
$^{\rm (24)}$\cite{klo},
$^{\rm (25)}$\cite{and}
$^{\rm (26)}$\cite{torii},
$^{\rm (27)}$\cite{math},
$^{\rm (28)}$\cite{jakobsson},
$^{\rm (29)}$\cite{rau},
$^{\rm (30)}$\cite{ay},
$^{\rm (31)}$\cite{malesani031203},
$^{\rm (32)}$\cite{fox},
$^{\rm (33)}$\cite{hu},
$^{\rm (34)}$\cite{terada},
$^{\rm (35)}$\cite{pri2},
$^{\rm (36)}$computed from \cite{costa},
$^{\rm (37)}$\cite{sta2},
$^{\rm (38)}$\cite{gcn2876},
$^{\rm (39)}$\cite{gcn2884},
$^{\rm (40)}$\cite{ber468},
$^{\rm (41)}$\cite{ber107},
$^{\rm (42)}$\cite{gcn3110},
$^{\rm (43)}$\cite{gcn3121},
$^{\rm (44)}$\cite{gcn3171},
$^{\rm (45)}$\cite{gcn3262},
$^{\rm (46)}$\cite{gcn3363},
$^{\rm (47)}$\cite{gcn3442},
$^{\rm (48)}$\cite{gcn3403},
$^{\rm (49)}$\cite{gcn3375},
$^{\rm (50)}$\cite{gcn3372},
$^{\rm (51)}$\cite{gcn3488},
$^{\rm (52)}$\cite{gcn3506},
$^{\rm (53)}$\cite{gcn3549},
$^{\rm (54)}$\cite{gcn3511},
$^{\rm (55)}$\cite{gcn3531},
$^{\rm (56)}$\cite{gcn3530},
\end{flushleft}
\end{table*}

\begin{figure*}
\centering
\resizebox{\hsize}{!}{\includegraphics[width=\columnwidth]{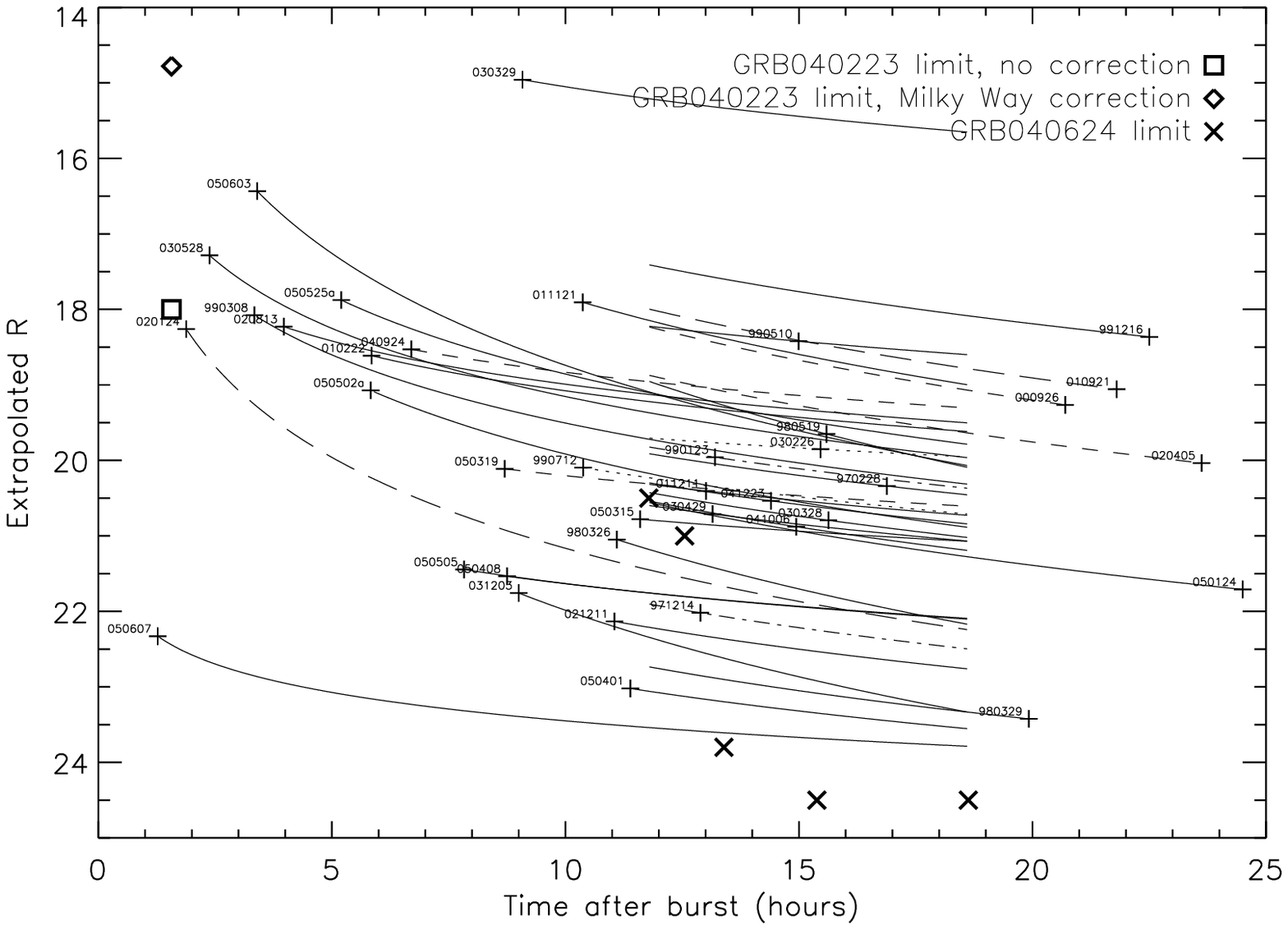}}
\caption{Light curves of the afterglows listed in
Table~\ref{tab:afterglow} with magnitudes extrapolated to
the $R$ band (when necessary) after correction for Galactic
absorption. The cross is placed at the time of observation.  For
clarity, different line styles are used to indicate the temporal
power-law decay extrapolations, which only cover our
observation epochs. The diamond and the square indicate the
magnitude limits for \object{GRB\,040223} with and without
correction for Galactic extinction, respectively, at
2-$\sigma$ (\cite{Gomb04}). The crosses indicate the magnitude
limits for \object{GRB\,040624}. The first two come from the GCNs
(\cite{Picc04,Goros04}). The last three points are our 3-$\sigma$
limits. } \label{fig:dark:040223:r}
\end{figure*}

\begin{figure*}
\centering
\resizebox{\hsize}{!}{\includegraphics[width=\columnwidth]{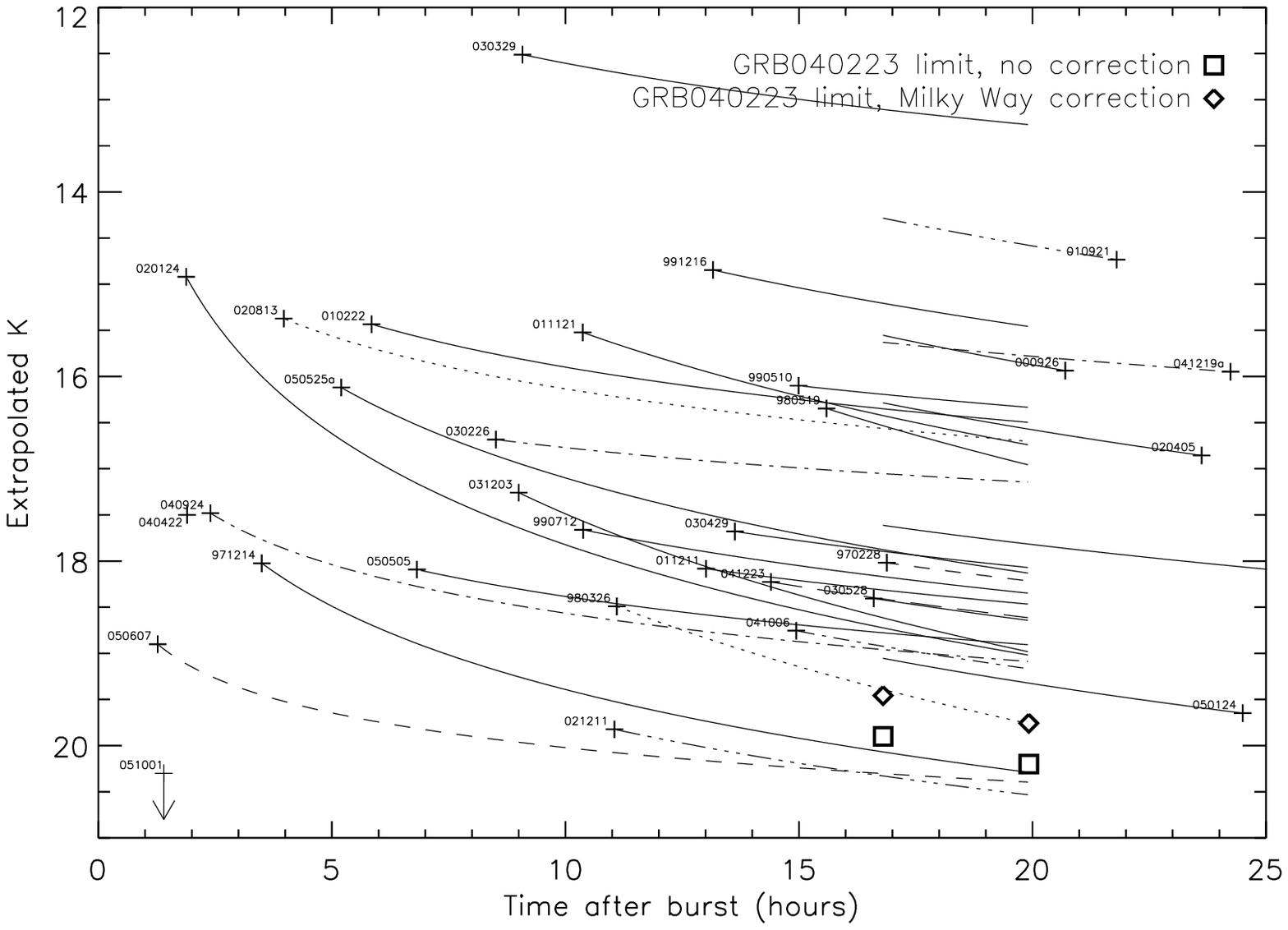}}
\caption{Light curves of the afterglows listed in
Table~\ref{tab:afterglow} with magnitudes extrapolated to
the $K_{\rm s}$ band (when necessary) after correction for
Galactic absorption. The cross is placed at the time of
observation. For clarity, different line styles are used to
indicate the temporal power-law decay extrapolations, which
only cover our observation epochs. The diamonds and the squares
indicate our 3-$\sigma$ magnitude limits with and without
correction from Galactic extinction, respectively. The $K_{\rm
s}$ magnitude for \object{GRB\,040422} (\cite{filliatre}) and the
limit for \object{GRB\,051001} (\cite{gcn4053}) are also
reported.} \label{fig:dark:040223:k}
\end{figure*}

We compared our magnitude limits for
\object{GRB\,040223} and \object{GRB\,040624} with the magnitudes
of observed afterglows that are listed in
Table~\ref{tab:afterglow} selected according to the following
criteria:
\begin{itemize}
\item a photometric measurement of the optical counterpart
was made in the first 30 hours in good observing
conditions. This timing constraint had two motivations:
first, to avoid a bias towards bright afterglows that late 
observations tend to select (as faint afterglows are then more
likely to be below the detection limit); second, to avoid
introducing a large error in extrapolating to the epochs of our
observations;
\item temporal index $\alpha$ is available to extrapolate to the first epoch of our
observations of \object{GRB\,040223} assuming that the light curve
can be modelled by $F(t,\nu)\propto t^{-\alpha}\nu^{-\beta}$; in
case of a temporal break, only the index valid for our epochs of
observations is retained;
\item the $R$ or $K_{\rm s}$ magnitude is given, with the delay preferably close to our epochs of observation; the
spectral index $\beta$, if given, is used to extrapolate from $R$
to $K_{\rm s}$ when necessary for comparisons with the
$K_{\rm s}$ band, and vice-versa.
\end{itemize}
\indent Adopting the spectral index, we extrapolate the
magnitudes to the $K_{\rm s}$ band and correct for absorption in
the Galaxy (yielding $A_R>1$ for \object{GRB\,991216},
\object{GRB\,011121},  \object{GRB\,030528}, \object{GRB\,031203}
and \object{GRB\,041219A}), using the $E(B-V)$ values from the
FIRAS maps, corrected according to \cite{dutra}. { Using the
temporal index, we then extrapolate these computed magnitudes to
the}  epochs of observation of \object{GRB\,040223} and
\object{GRB\,040624}. The temporal curve may deviate from a
power-law behaviour
 (e.g. Tagliaferri et al.~2005a): however in the present case we compare magnitudes far from the early times.
 Galactic extinction is subject to inaccuracy, especially at low Galactic latitudes. However, the afterglow
 of \object{GRB\,011121} is the only case for which we made an extrapolation to the $K$ band with $A_R(\rm Galaxy)>1$.
 The lowest value is that of  \object{GRB\,021211}, which is $A_R(\rm Galaxy)=0.067$.
 Estimating the extinction within the host galaxy for the afterglows of Table~\ref{tab:afterglow} is
beyond the scope of this paper. However, the
 indications are that the reddening might be low. In fact in the rest frame of the host, $A_V(\rm host)=0.21\pm 0.12$
 for \object{GRB\,971214} and consistent with zero for \object{GRB\,980519}, \object{GRB\,990123}, and \object{GRB\,990510}
 (\cite{stratta}). For \object{GRB\,020813}, \cite{savaglio} infer a relatively high extinction of
 $A_V(\rm host)\simeq 0.4$, but with weak dependence on wavelength, and therefore little impact on the power-law
 shape of the spectrum. This suggests that the extinction within the
   host is small, at least for the afterglows listed in Table~\ref{tab:afterglow} and observed in the $R$ band.\\
\indent The results are presented in Figs~\ref{fig:dark:040223:r} and \ref{fig:dark:040223:k} in $R$ and $K_{\rm s}$ bands.
For each epoch we show the observed magnitude limit. For \object{GRB\,040223} we report both the absorbed and unabsorbed
value  (the absorption for \object{GRB\,040624} can be neglected here).\\
\indent For \object{GRB\,040624}, our deep 3-$\sigma$ limits reported in Fig.~\ref{fig:dark:040223:r} indicate that we are
 facing a faint afterglow, even fainter than the Swift afterglow of \object{GRB\,050607}. Indeed, gaussian fits
 of the distribution of the extrapolated magnitudes shown in Fig.~\ref{fig:dark:040223:r} at our observation epochs
 indicate that the afterglow of \object{GRB\,040624} is fainter by at least $2.7\,\sigma$. As the sample of bursts
 in Fig.~\ref{fig:dark:040223:r} is constructed to limit the bias towards bright afterglows, this indicates that the
 afterglow of \object{GRB\,040624} is fainter than normal. This is all the more interesting because this burst was
 very well placed for an observational campaign: near the celestial equator and at high Galactic latitude. We can then
 consider \object{GRB\,040624} as a very good example of a  dark burst. Given the lack of observations at earlier
 epochs and at other wavelengths, the causes of its darkness cannot be properly ascertained.\\
\indent For \object{GRB\,040223}, our observations in the $r$
band are not so deep due to the location of the GRB towards
the Galactic plane and the fairly long delay time for the $K_{\rm
s}$ measurements. However, our limits (and especially the first
epoch one, which is the more stringent) indicate that the burst is
faint in the $K_{\rm s}$ band, as only a few GRBs have
extrapolated magnitudes similar to our limit (i.e.
\object{GRB\,971214}, \object{GRB\,021211} and
\object{GRB\,050607}). In the $J$ and $H$ band, because of the
Galactic absorption, our magnitude limits are less
restrictive. Note that our group has already detected the
faint afterglow of \object{GRB\,040422} with the VLT/ISAAC
(\cite{filliatre}), with an observed magnitude $K_{\rm
s}=18.0\pm0.1$ and a Galactic absorption of  $A_K=0.5$
similar to that for \object{GRB\,040223}. However, 
that observation was much quicker, 1.9 hours after the burst.
Assuming a temporal decay index of 1, the afterglow of
\object{GRB\,040422} would have been $K_{\rm s}=20.3$ at our
first epoch of observation of \object{GRB\,040223} and
much fainter than our detection limit. The case of
\object{GRB\,051001}, with a $3\,\sigma$ magnitude limit of
$K_{\rm s}=20.3$ only 1.40 hours after the 
trigger, (\cite{gcn4053}) is even more extreme, and demonstrates
that observing within the first few hours is mandatory in
the quest for faint afterglows.

\subsection{\object{GRB\,040223}: from the X-ray to the NIR afterglow}

\begin{figure}
\centering
\resizebox{\hsize}{!}{\includegraphics[width=\columnwidth,angle=270]{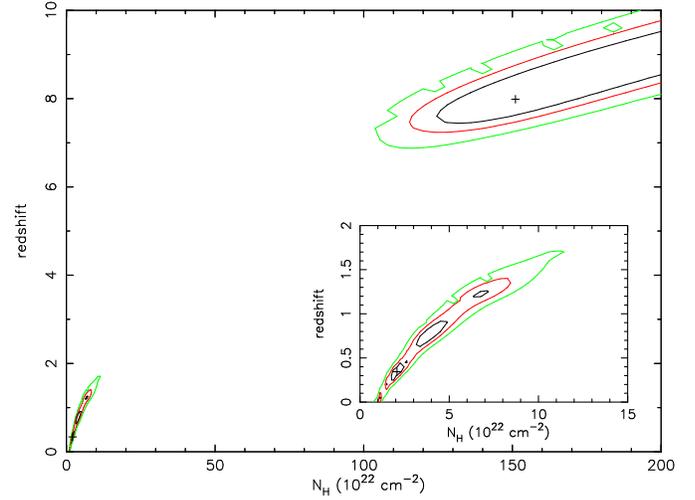}}
\caption{{ For \object{GRB\,040223}}, confidence contours (at
68\%, 90\% and 99\% level for two parameters of interest) for the
redshift $z$ and the column density $N_{\rm H,z}$ of the
redshifted absorber, derived from the fit to the EPIC spectra. The
inner plot is an enlargement of the low-$z$ region. }
\label{confcont}
\end{figure}

The power-law indices of the X-ray afterglow of
\object{GRB\,040223}  are $\alpha_{\rm X}=0.9\pm0.1$ and
$\beta_{\rm X}=1.60\pm0.15$ and are not compatible with the
closure relation in the form $\alpha+b\beta+c=0$ for isotropic
expansion into a homogenous medium (\cite{sari}), isotropic
expansion into a wind-stratified medium (\cite{chevalier}) or a
collimated expansion into a homogenous or wind-stratified medium
(\cite{sari2}), the cooling frequency being bluewards or redwards
the X-ray domain. This fact can be seen in
Figure 2 of \cite{piro}, with the other, less compelling,
case of \object{GRB\,980703}.\\
\indent The extremely high hydrogen column density derived from
the spectrum of the X-ray afterglow is significantly higher than
the estimated Galactic hydrogen column density. Assuming for
simplicity that there is no significant absorption (in X-rays,
optical and NIR) between our Galaxy and the GRB host, this
strongly suggests that intrinsic absorption, in the
restframe of the GRB, is present at a significant level. The
correction needed in the optical/NIR bands to take this
absorption into account, depends mainly on two unknown
factors. The first one is the redshift of the GRB, which is
required to relate the observed wavelength to the absorbed
emitted wavelength, and also to estimate properly the hydrogen
column density in the host galaxy. It is possible to use the
spectrum of the X-ray afterglow to put constraints on the redshift
of the burst and the intrinsic column density. For that, a
redshifted model for the interstellar absorption, including a
series of photoelectric absorption edges is fitted to the X-ray
spectrum assuming solar system abundances. The resulting best fit
value for the redshift is $z\sim0.35$, for a corresponding
intrinsic column density $N_{\rm H}^{\rm host}
(z)\sim2\times10^{22}\,\rm cm^{-2}$ (in addition to the
previously quoted value $N_{\rm H}^{\rm
Galaxy}=(6.6\pm2)\times10^{21}\,\rm cm^{-2}$), but with no
significant improvement in the quality of the fit with
respect to the simple absorbed
 power-law model
 ($\chi^2_{\nu}=0.93$, 113 d.o.f.). As the photoelectric cross-section scales roughly as $E^{\sim -2.6}$, we
 have, to first approximation  $N_{\rm H}^{\rm host} (z)=N_{\rm H}^{\rm host} (z=0)(1+z)^{2.6}$,
 where $N_{\rm H}^{\rm host} (z=0)=N_{\rm H}^{\rm measured}-N_{\rm
H}^{\rm Galaxy}$ (\cite{stratta}). Indeed, the redshift and
the column
 density parameters in the fit are degenerate, as may be
 seen from the contour plot of Fig.~\ref{confcont}. In any case, 
the data allows us to estimate that the gas column is at least
$N_{\rm H}^{\rm host} (z=0)=10^{22}$ cm$^{-2}$, while the redshift
is no larger than 1.7 ($N_{\rm H}^{\rm host} (z)<1.1\times10^{23}$
cm$^{-2}$), at 90\% confidence level for 2 parameters of interest.
For $z>1.7$ the fit turns out to be unacceptable since we try to
fit the \element{Fe} edge in the portion of the spectrum which has
larger statistics (below 2.5 keV) and which does not exhibit an
absorption edge. Note however that the fit is good again for
$z>7$, for which there is no absorption edge in the 1-4 keV
range. Hence, the X-ray data constrain the
redshift to be $0<z<1.7$, or $z>7$, assuming solar abundances. \\
\indent The second factor is the relationship between the hydrogen
column density and the absorption law that is valid in the
optical/NIR domain. We will consider two hypotheses: first, the
host galaxy behaves like our Galaxy, and the relationships
of \cite{predehl}
 and \cite{cardelli} with $R_{V}=3.1$ still hold; second, the host galaxy behaves as the Small Magellanic Clouds, with
 $N_{\rm H}/A_V=1.6\cdot10^{22}\,\rm cm^{-2}$ (\cite{weingartner}) and use the extinction law fitted by \cite{pei}
 with $R_V=2.93$. This hypothesis is a reasonable choice for the host, as far as absorption properties are concerned
 (\cite{stratta}).\\
\indent With these two hypotheses, and assuming that $N_{\rm H}^{\rm host} (z)=N_{\rm H}^{\rm host} (z=0)(1+z)^{2.6}$,
we extrapolated the luminosity of the X-ray afterglow to the $r$ to $K$ bands, and compared the result with our
observational limits. Assuming a synchrotron emission for the afterglow, we considered two cases:
\begin{enumerate}
\item both the photon and time indices derived in the X-ray are valid in the $r$ to $K$ bands, at the epochs of
observations; the degree of freedom here is the redshift;
\item the cooling break frequency happens to lie between the X-ray domain and the optical/NIR domain at the {
observation epochs} in the optical/NIR, with two { different spectral distributions} (\cite{sari}):
\begin{enumerate}
\item the fast cooling case, for which $\beta_{\rm opt,\,NIR}=0.5$;
\item the slow cooling case, for which $\beta_{\rm opt,\,NIR}=\beta_{\rm X}-0.5$.
\end{enumerate}
\end{enumerate}
The extrapolated magnitudes are listed in
Tables~\ref{tab:xray:extrapol:mw} and
\ref{tab:xray:extrapol:smc}, where the error bars are computed by
simulations taking into account the uncertainties in the
parameters (the X-ray flux, $\beta_{\rm X}$, $\alpha_{\rm X}$,
$N_{\rm H}^{\rm measured}$ and  $N_{\rm H}^{\rm Galaxy}$). In
all cases, the most stringent constraints are provided by
the observations in the $K_{\rm s}$ band. We can also assume that
a temporal break occurred between the epoch of the X-ray 
observations, and the epoch of the $J$, $H$, $K_{\rm s}$ 
observations. This case can combine with the previous ones,
independently of the modelling of the absorption in the host
galaxy and of the redshift. However, even assuming { that the
temporal index steepened to $\alpha=2$ just at the end of
the X-ray observations, the extrapolated NIR afterglow would be
brighter by less than 1 magnitude, which is smaller than the
error bars reported in Tables~\ref{tab:xray:extrapol:mw} and
\ref{tab:xray:extrapol:smc}.} We therefore do not consider further
the possibility
of a temporal break.\\
\indent These results are compatible with the hypothesis that the synchrotron model is valid and that the X-ray and
optical/NIR afterglow come from the same component, although the hypothesis that the afterglows are due to two different
components cannot be excluded either. Indeed, the extrapolations are compatible with the observational limits in the
following cases:
\begin{enumerate}
\item $z\gse 0.9$ if the host is similar to the Galaxy, $z\gse 2.1$  if the host is similar to the SMC;
\item with a cooling break,
\begin{enumerate}
\item with a fast cooling, for  a cooling break frequency $\nu_c>7.5\cdot10^{16}\,\rm Hz$ ($\lambda_{c}<0.01\,\mu\rm m$)
if the host is similar to the Galaxy;  for a cooling break
frequency at the lower edge of the X-ray band ($E_{c}=0.5\,\rm
keV$, i.e. $\nu_c=1.2\cdot10^{17}\,\rm Hz$), or for
$\nu_c>7.5\cdot10^{16}\,\rm Hz$ for $z\gse 0.8$ if the host is
similar to the SMC;
\item with a slow cooling, for $\nu_c=1.2\cdot10^{17}\,\rm Hz$  and $z\gse 0.6$ if the host is similar to the Galaxy,
$z\gse 1.8$ if the host is similar to the SMC.
\end{enumerate}
\end{enumerate}
However the possibility that these two GRBs are intrinsically
sub-luminous with faint afterglows cannot be excluded. Redshifts
above 0.9 are common in GRBs and case 1, because it does not
involve fine tuning between the epochs of observations and a break
in the spectrum, is the simplest explanation. In conclusion, our
magnitude limits are compatible with a simple extrapolation of the
properties of the X-ray afterglow for an acceptable redshift
range, independent of the host galaxy.

 \begin{table*}
\centering\caption{The magnitudes of the optical/NIR
afterglow of \object{GRB\,040223} predicted by extrapolation of
the X-ray afterglow, assuming that the host is similar to the 
Galaxy. The observed limits are reported for comparison.}
\label{tab:xray:extrapol:mw}
\begin{tabular}{llcccccccl}
\hline
&$r$&$J$, $1^{\rm st}$ epoch&$J$, $2^{\rm nd}$ epoch&$H$, $1^{\rm st}$ epoch&$H$, $2^{\rm nd}$ epoch&$K_{\rm s}$, $1^{\rm st}$ epoch&$K_{\rm s}$, $2^{\rm nd}$ epoch\\
\hline
Observed limits&18&20.9&21.1&19.8&20.4&19.9&20.2\\
\hline
Simple extrapolation&&&&&&&\\
$z=0$&$19.5\pm2$&$16.0\pm2$&$16.2\pm2$&$14.1\pm2$&$14.3\pm2$&$12.5\pm2$&$12.6\pm2$\\
$z=0.9$&$58\pm11$&$37\pm6$&$37\pm6$&$27\pm4$&$27\pm4$&$20.7\pm3$&$20.9\pm3$\\
\hline
Cooling break, fast cooling&&&&&&&\\
$z=0$, $\nu_c=7.5\cdot10^{16}\,\rm Hz$&$25.6\pm1$&$22.8\pm1$&$23.0\pm1$&$21.2\pm1$&$21.4\pm1$&$20.0\pm1$&$20.1\pm1$\\
$z=0$, $\nu_c=1.2\cdot10^{17}\,\rm Hz$&$26.2\pm1$&$23.4\pm0.5$&$23.6\pm0.5$&$21.9\pm0.4$&$22.0\pm0.4$&$20.6\pm0.3$&$20.7\pm0.3$\\
\hline
Cooling break, slow cooling&&&&&&&\\
$z=0$, $\nu_c=1.2\cdot10^{17}\,\rm Hz$&$22.5\pm2$&$19.3\pm2$&$19.5\pm2$&$17.6\pm2$&$17.8\pm2$&$16.1\pm2$&$16.3\pm2$\\
$z=0.6$, $\nu_c=1.2\cdot10^{17}\,\rm Hz$&$42\pm6$&$29\pm3$&$29\pm3$&$23.6\pm2$&$23.7\pm2$&$19.9\pm2$&$20.1\pm2$\\
\hline
\end{tabular}
\end{table*}

 \begin{table*}
\centering\caption{The magnitudes of the optical/NIR
afterglow of \object{GRB\,040223} predicted by extrapolation of
the X-ray afterglow, assuming that the host is similar to the SMC.
The observed limits are reported for comparison.}
\label{tab:xray:extrapol:smc}
\begin{tabular}{llccccccc}
\hline
&$r$&$J$, $1^{\rm st}$ epoch&$J$, $2^{\rm nd}$ epoch&$H$, $1^{\rm st}$ epoch&$H$, $2^{\rm nd}$ epoch&$K_{\rm s}$, $1^{\rm st}$ epoch&$K_{\rm s}$, $2^{\rm nd}$ epoch\\
\hline
Observed limits&18&20.9&21.1&19.8&20.4&19.9&20.2\\
\hline
Simple extrapolation&&&&&&&\\
$z=0$&$15.8\pm2$&$14.6\pm2$&$14.7\pm2$&$13.2\pm2$&$13.3\pm2$&$11.9\pm2$&$12.0\pm2$\\
$z=2.1$&$50\pm9$&$32\pm5$&$33\pm5$&$26\pm4$&$26\pm4$&$20.4\pm3$&$20.6\pm3$\\
\hline
Cooling break, fast cooling&&&&&&&\\
$z=0.8$, $\nu_c=7.5\cdot10^{16}\,\rm Hz$&$25.6\pm1$&$23.2\pm0.6$&$23.3\pm0.6$&$21.5\pm0.4$&$21.6\pm0.4$&$20.1\pm0.4$&$20.2\pm0.4$\\
$z=0$, $\nu_c=1.2\cdot10^{17}\,\rm Hz$&$22.5\pm0.3$&$22.0\pm0.3$&$22.2\pm0.3$&$20.9\pm0.3$&$21.1\pm0.3$&$20.0\pm0.3$&$20.2\pm0.3$\\
\hline
Cooling break, slow cooling&&&&&&&\\
$z=0$, $\nu_c=1.2\cdot10^{17}\,\rm Hz$&$18.8\pm1$&$17.9\pm1$&$18.1\pm1$&$16.7\pm1$&$16.8\pm1$&$15.5\pm1$&$15.7\pm1$\\
$z=1.8$, $\nu_c=1.2\cdot10^{17}\,\rm Hz$&$39\pm6$&$28.6\pm3$&$28.7\pm3$&$24.1\pm2$&$24.3\pm2$&$20.3\pm2$&$20.5\pm2$\\
\hline
\end{tabular}
\end{table*}

 \section{Conclusions}

Our main results on \object{GRB\,040223} and
\object{GRB\,040624} are as follows:
\begin{itemize}
\item These two bursts are among the weakest and longest detected by \textit{INTEGRAL}. The $\gamma$-ray spectra are
well fit by power-laws with steep spectral indices 
suggesting that both are X-ray rich GRBs.
\item For \object{GRB\,040223}, an X-ray afterglow was detected by \textit{XMM-Newton}. The spectral properties
indicate an extremely high column density of
$(1.60\pm0.15)\times10^{22}\,\rm cm^{-2}$, much higher than the
expected contribution from the Galaxy. This result
suggests significant absorption in the host galaxy of the burst.
\item The magnitude limits we obtained in the NIR for \object{GRB\,040223} are compatible with a simple extrapolation of
the properties of the X-ray afterglow.
\item For \object{GRB\,040624}, we obtained magnitude limits in the optical that are fainter than the very faint end of
the distribution of the magnitudes of a compilation of 39 promptly observed counterparts.
\end{itemize}
In conclusion these two GRBs can be classified as bursts
with a dark afterglow. For \object{GRB\,040223}, the cause of
darkness seems to be the intrinsic absorption. We remark that a
very high dust extinction ($A_{V}\sim30-50$) has also been
proposed as a possible cause of the discrepancy (a factor 3-10)
between the observed NIR rate of core-collapse supernovae and the
rate estimated from the far-IR luminosity of the parent galaxies
(\cite{mannucci,maiolino}). Given the relationships between long
GRBs and core-collapse supernovae (e.g. \cite{dellavalle}), we may
be led to the conclusion that a similar effect is at play also
in this case and that most "dark" bursts are in fact highly
extincted "normal" GRBs.\\
\indent The increasing use of large diameter telescopes, with
instruments operating in the infrared, involved in a comprehensive
observational programme with reduced delay may increase the
number of similar bursts. The magnitude of \object{GRB\,040422}
(\cite{filliatre}) and more dramatically the limit on
\object{GRB\,051001} (\cite{gcn4053})
stress the  clear fact that, when looking for faint afterglows, observing within the first few hours is mandatory. \\

 \begin{acknowledgements}
We are very grateful to the ESO staff at Paranal for carefully
performing all our observations, and for many useful suggestions. This research is
supported by the Italian Space Agency (ASI), and is part of the GRB
activity related to the Swift mission (contract ASI/I/R/390/02).
\end{acknowledgements}

 \end{document}